\documentclass[10pt]{article}

\usepackage{color}
\usepackage{graphicx}
\usepackage{multicol}
\usepackage{verbatim}
\usepackage{authblk}

\title{Some new results for the GHWS model}
\author[]{Leonardo Reyes}
\affil[]{{\small Laboratorio de Din\'amica no-lineal y Sistemas Complejos, Centro de F\'isica, Instituto Venezolano de Investigaciones Cient\'ificas (IVIC), Rep\'ublica Bolivariana de Venezuela.\\
{\tiny leonardoivanrc@gmail.com}}}

\begin{document}

\maketitle

\begin{abstract}
Here we outline some new results for the GHWS model which points to a discretization of parameter space into well differentiated collective dynamic states.
We argue this can lead to basic processes in parameter space, starting with minimum modelling ingredients: a complex network with a disorder
parameter and an excitable dynamics (cellular automata) on it. 
We relate {\it energy loss} and {\it dynamics} for processes in parameter space with constant fluctuations of activity.
\end{abstract}

\vspace*{10px}
\rightline{{\it `dynamics ﬁrst and consolidation to symbols later'}}
\vspace*{20px}

In reference \cite{ghws} numerical results for the GHWS model were presented. One of the main results was that we can change the collective state of the system
by changing network disorder alone, at constant coupling $\sigma=rK$, where $K$ is the average coordination number in the WS network and $r$ is the transmission coefficient in the GH automaton \cite{Chialvo} (see also \cite{Gross} for some related ideas). 
We reproduce equation $(2)$ from reference \cite{ghws} here:

\begin{equation}\label{eq2old}
\sigma_c=\Delta_\sigma\exp[-(x/\tau)^\beta]+\sigma_{c1}
\end{equation}

which summarizes numerical results for the frontier between extinct and active collective states. 
That frontier in \cite{ghws} was obtained by considering as extinct states those for which, after $T$ steps, the activity was zero ($F=0$) for all $M$ generated samples which differ in their random initial condition and rewiring. 
Equation (\ref{eq2old}) is a stretched exponential (with exponent $\beta$) in the rewiring odd $x=p/(1-p)$, where $p$ is the rewiring probability in the WS network.

\begin{figure}[ht]
\centering
\includegraphics[width=0.7\textwidth]{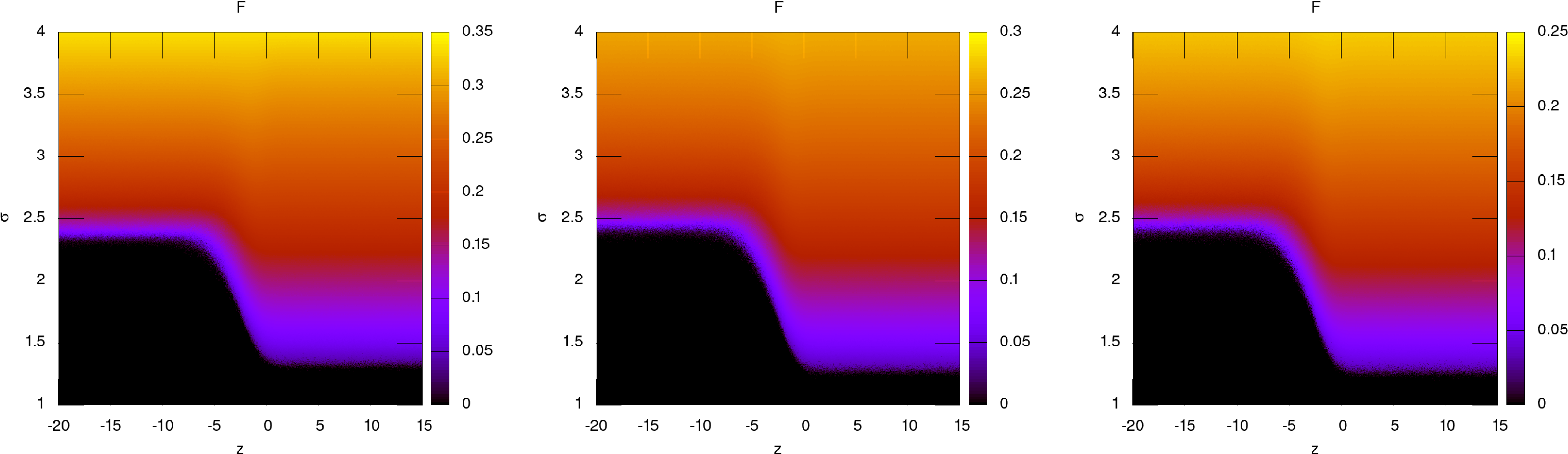}
\caption{Activity $F(z,\sigma)$ for $K=4$ (left), $K=6$ (center), $K=8$ (right). $N=1000$. In this figure and also on figures \ref{figKs} and \ref{figK6N} we plot $1000\times 1000$ points
in parameter space $(z,\sigma)$, each of them obtained from a single run. We depict in black the absorbing state.}
\label{figFKs}
\end{figure}

Further numerical results for the activity $F(z,\sigma)$ in parameter space can be seen in figure \ref{figFKs}, with disorder $z=\log x=\log[p/(1-p)]$ and with the absorbing state depicted in black. 
We have confirmed that the following simplification to eq. (\ref{eq2old}) holds:

\begin{equation}\label{eq2new}
\sigma_c=\Delta_\sigma\left(1+\exp[-(x/\tau)^\beta]\right)
\end{equation}

that is: $\sigma_{c1}\approx\Delta_\sigma$, with $\Delta_\sigma\approx 1.2$ from our numerical experiments, for some $K$ and $N$ (see below). 

\begin{figure}[ht]
\centering
\includegraphics[width=0.7\textwidth]{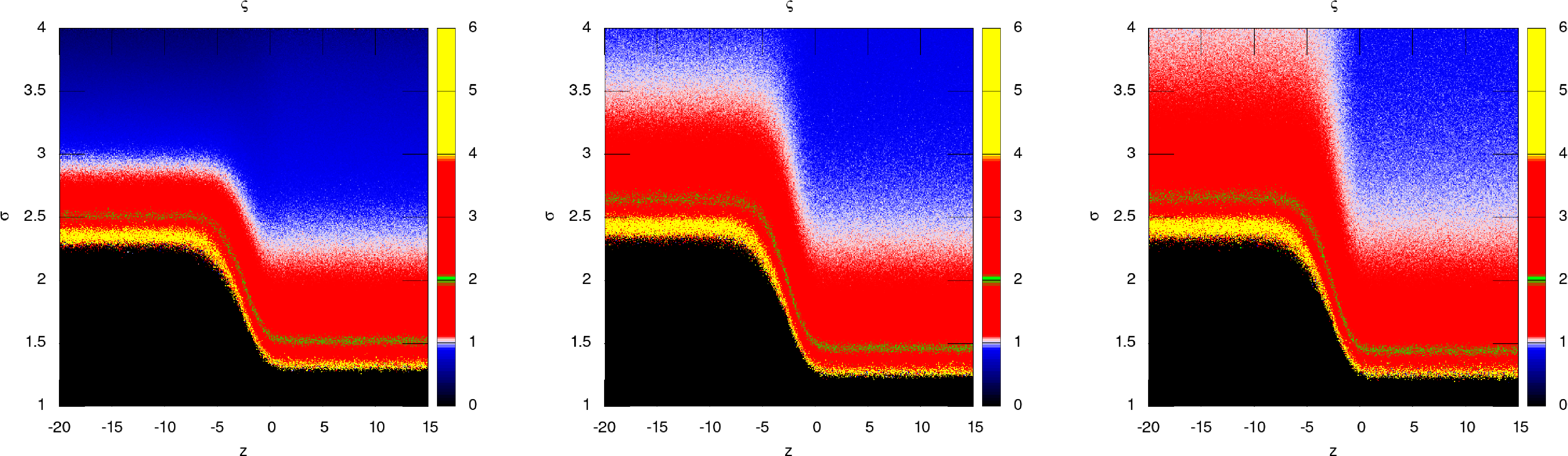}
\caption{Normalized fluctuations of activity $\varsigma(z,\sigma)$ for $K=4$ (left), $K=6$ (center), $K=8$ (right). Note in green the reference value $\varsigma=2$, which is close
to the extinct-active boundary in parameter space. After a transient of $T=1000$ time steps we considered the next $1000$ steps for obtaining the fluctuations. $N=1000$.}
\label{figKs}
\end{figure}

\begin{figure}[ht]
\centering
\includegraphics[width=0.7\textwidth]{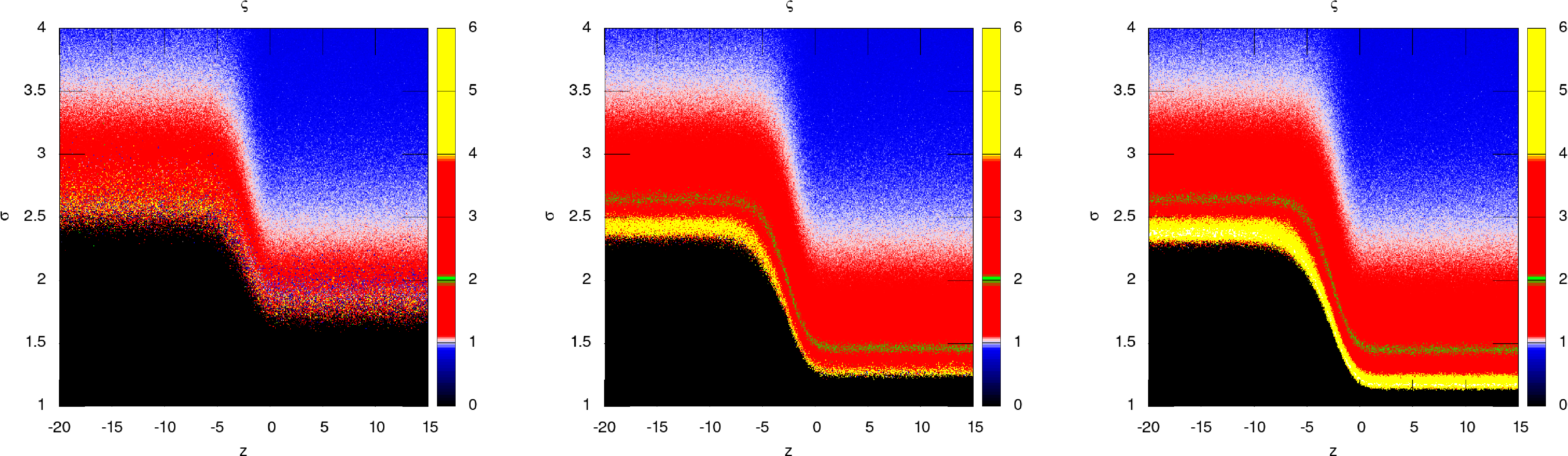}
\caption{Normalized fluctuations of activity $\varsigma(z,\sigma)$ for $K=6$ with $N=100$ (left), $N=1000$ (center), $N=10000$ (right). Note in green the reference value $\varsigma=2$, which is close
to the extinct-active boundary in parameter space. }
\label{figK6N}
\end{figure}

Even more relevant than $F$ for a possible connection with experiments should be the (normalized) fluctuations of activity $\varsigma$ \cite{Chandler}:

\begin{equation}
\varsigma=\frac{N}{\langle F\rangle(1-\langle F\rangle)}\left[\langle F^2\rangle -\langle F\rangle^2\right]
\end{equation}

with $\varsigma\approx 1$ implying a near ideal gas like dynamics (no correlations of activity) \cite{Chandler}. 
$\varsigma(z,\sigma)$ can be appreciated in figures \ref{figKs} and \ref{figK6N} for some values of $K$ and $N$, where $N$ is the number of nodes in the network.
For $K=6$ and $N=1000$ we can see the further relevance of the quantity $\Delta_\sigma$, where the curve $\varsigma=1$ is obtained by displacing the extinct-active frontier as 
$\sigma\rightarrow\sigma+\Delta_\sigma$ in the whole
range of disorder $z$. The curve $\varsigma=1$ (for $K=6$) in parameter space in fact provides a better method for measuring $\Delta_\sigma$, which can be used to locate back the extinct-active
frontier. 

Thus, the quantity $\Delta_\sigma$ is not only a disorder-dependent activation threshold, 
but provides further structure in parameter space in connection with a dynamic state for which we have a clear physical interpretation ($\varsigma=1$).

We can see from figures \ref{figFKs}, \ref{figKs} and \ref{figK6N} that we can talk about of two phases in the GHWS model, for which $\partial\varsigma/\partial z=0$ (and $\partial F/\partial z=0$):
an ordered phase (say $z<-10$) and a disordered phase (say $z>4$). Fluctuations and patterns of activity for both phases can be seen in figures \ref{varsigma12} and \ref{patterns}.
Note the different way in which large correlations of activity manifest themselves in the ordered and disordered phases near the extinct-active frontier 
(fluctuations $\Leftrightarrow$ correlations \cite{Chandler}).

Thus, at least for $K=6$, we obtain characteristic and differentiated dynamic states for couplings $\sigma$ near:

\begin{equation}\label{sigma-ms}
\sigma_{ms}=\Delta_\sigma(1+s+m)
\end{equation}

where $s=0,1$ and $m=0,1,2,3,\ldots$ are integer numbers. For the ordered phase $s=1$, and $s=0$ for the disordered phase. 
For $(m,s)=(0,0)$ we get a dynamic state near the extinct-active frontier in the disordered phase.
For $(m,s)=(0,1)$ we get a dynamic state near the extinct-active frontier in the ordered phase. For $(m,s)=(1,0)$ and $(m,s)=(1,1)$ we obtain a state with $\varsigma=1$.
If $\varsigma<1$ ($m>1$) then we have negative average correlations of activity, and we have states with random-like patterns of activity, with high values of activity (see figure \ref{figFKs}),
and with a peak in the power spectra of activity at frequency $1/3$ (at least for small $K$): the microscopic natural frequency for each {\it neuron} with three states surfaces to 
macroscopic activity, see figure \ref{fou}. That {\it microscopic natural frequency} equal to $1/3$ is certainly the case in the deterministic limit $r\rightarrow 1$.

\begin{figure}[ht]
\centering
\includegraphics[width=0.8\textwidth]{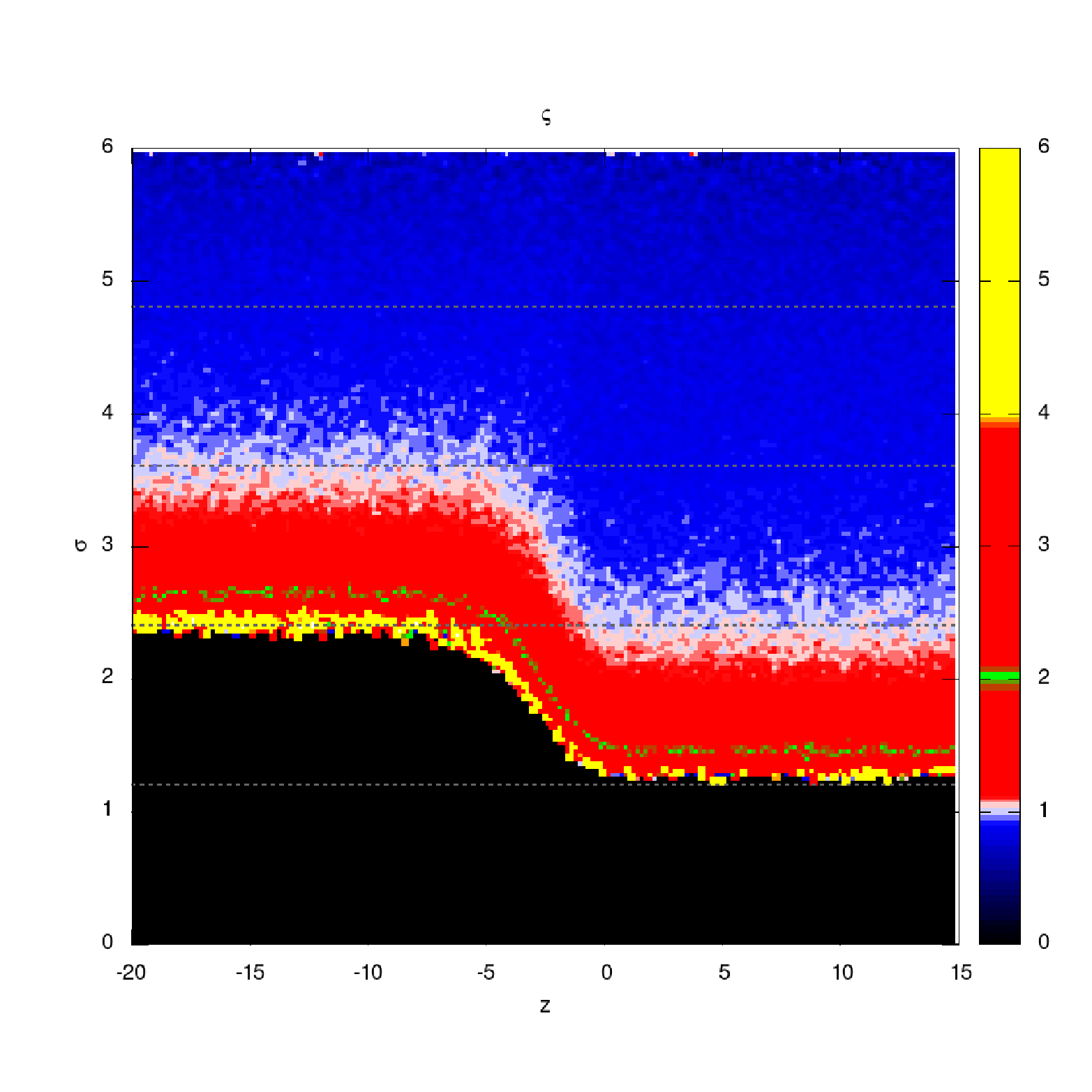}
\caption{Discretization of parameter space for $K=6$ and $N=1000$. We show $200\times 200$ points. In dashed horizontal lines we show $\sigma$ in multiples of $\Delta_\sigma=1.2$.}
\label{deltasigma}
\end{figure}

By looking at figures \ref{figKs} and \ref{figK6N} it seems that how good is equation (\ref{sigma-ms}) 
for describing our numerical results is a function of average coordination number $K$ and
number of nodes $N$. Around $K=6$ and $N=1000$ it appears to be a good approximation. 
Thus, for certain $K$ and $N$, we obtain a natural discretization of parameter space, a {\it consolidation to symbols} \cite{Kaneko}, see figure \ref{deltasigma}.
It would be natural now to consider processes in parameter space, between these reference system's states labeled by integer numbers $(m,s)$.

From a mean field approximation we can obtain $\sigma_c=1$. It is natural then to write $\Delta_\sigma=1+\triangleright$, with the quantity $\triangleright$
associated to correlations of activity. We could rewrite equation (\ref{sigma-ms}) as

\begin{equation}\label{new-sigma-ms}
\sigma_{ms}=(1+\triangleright)(1+s+m)
\end{equation}

See new figure \ref{new} for a numerical verification of this. At minimums we calculate $\Delta_\sigma=1+\triangleright$, with $\Delta_\sigma$ obtained from the value of $\sigma$
at the transition in the disordered phase (see also the caption of figure \ref{new}). Numerically, we obtain de relation:

\begin{equation}\label{triangulo}
\triangleright/a=K+c
\end{equation}

with the constant $a\approx 1/9.5$ : correlations proportional to coordination number. For each $K$ we obtain an $N$ that maximizes how cuadriculate ({\it discretization}) is parameter space.\\

It can be expected some relationship between $\triangleright$ and $\varsigma$, and may be that connection could be found in the behaviour of
$\varsigma(\sigma)$ around $\varsigma=1$ for the ordered and disordered phases, see figure \ref{varsigma12}.\\

***\\

We could get further guidance by trying to emulate a macroscopic analog of the microscopic dynamic rule (the GH automaton). This requires to identify the {\it excited}, {\it susceptible} and {\it refractory} macro-states, and then connect them by processes in parameter space.
Sigmoid functions are usually used when talking about {\it learning}. We could associate curves $\sigma_\varsigma(z)$, for given $\varsigma=constant$, with a learning process. This would lead us to associate {\it time} with an increase in $z$ \cite{ghws}.
One possibility then is to associate the susceptible macro-state with the very ordered networks of sigmoid $\sigma_\varsigma(z)$,
the excited macro-state for $z\sim z_c$ of $\sigma_\varsigma(z)$, and the refractory macro-state with the very disordered networks.
For modeling, we could invoke also random (or not) reinsertion processes.
Only for $z\sim z_c=\ln(\tau)$ we can change the collective state of the system at constant coupling $\sigma$ by changing network disorder alone. In figure \ref{var1} it is shown $\sigma_\varsigma(x)$ for $\varsigma\approx 1$. If $\sigma_\varsigma(x)$ is a stretched exponential function then $\sigma_\varsigma(z)$ is
a double exponential function. The larger the value of $\varsigma$ the closer to the collective extinct-active frontier (see figure \ref{deltasigma}). 
$\varsigma<1$ implies negative average correlations of activity (see figure \ref{fou}).\\

We can obtain the activity $F$ along $\sigma_\varsigma(x)$ as $x$ changes, for given $\varsigma$ (that is: $F_\varsigma(x)$). 
In figure \ref{figx3} we show the decay of activity as a function
of the odd of a rewiring event $x=p/(1-p)$ for three values of the normalized fluctuations of activity $\varsigma$, and they also are well described
with a stretched exponential function. In figure \ref{figx4} we show $\Delta_\sigma(\varsigma)$, $\Delta_F(\varsigma)$, the exponents $\beta_\sigma(\varsigma)$
and $\beta_F(\varsigma)$ and, most important, $\beta_F(\Delta_F)$.\\

$\beta_F(\Delta_F)$ is a relation between {\it energy lost} ($\Delta_F$) and {\it dynamics} ($\beta_F$) \cite{Elton} for a given $\sigma_\varsigma(x)$ process, 
a relation that was obtained through the quantity $\varsigma$.
These results points to consider processes $\sigma_\varsigma(x)$ as relaxation processes, in which connectivity is evolving from 
order to disorder. If we associate activity $F$ to {\it energy}, then $\varsigma(F)$ and $\int \varsigma dF$ are
very relevant quantities \cite{Chandler}. \\

What could be learned along such $\sigma_\varsigma(x)$ processes? It could be said that what could be learned is the $\beta$ parameter, which 
give us the optimal ({\it a la} PCA) direction for parameter fluctuations when $z\sim z_c$, which in turn would determine the exponent for the power law decay 
in the last stage of the process (very disordered networks, the refractory macro-state). \\

In figure \ref{profs} we show $F_\varsigma(z)$ for some $\varsigma<1$. 
At $\varsigma\sim0.82$ there's a {\it bump} in the profile, just before the decay in activity,
and at this value of $\varsigma$ $\beta\rightarrow 2$ (the gaussian, see the inset). Thus, we obtain that for $\varsigma\le0.82$ ($\beta\ge2$) there is an additional {\it cost} 
(an increase in activity) associated to those processes.

\newpage

\begin{figure}[ht]
\centering
\includegraphics[width=0.75\textwidth]{newfigv2.eps}
\caption{Normalized fluctuations of activity $\varsigma$ for ordered and disordered phases ($K=6$, $N=1000$). 
Note on the left the overlapping of $\varsigma(\sigma)$ for ordered and disordered phases once displaced by $\Delta_\sigma=1.2$ (for $\varsigma\in(0.9,2.3)$).
Also, on the right, we show the variation of the fraction of active samples after $T=1000$ time steps $\Pi$ with respect to the coupling parameter $\sigma$ as a function of $\sigma$.
At left and right we show averages over $M=1000$ samples. At center we show results from single runs.}
\label{varsigma12}
\end{figure}

\begin{figure}[ht]
\centering
\includegraphics[width=0.7\textwidth]{patterns.eps}
\caption{Patterns of activity. From right to left we get closer to the extinct-active frontier. 
{\it top}: disordered phase ($\sigma=1.3, 1.4, 1.5, 1.6$). {\it bottom}: ordered phase ($\sigma=2.4, 2.6, 2.8, 3.0$). 
 ($K=6$, $N=1000$). For these eight plots: random initial conditions only on $100$ nodes in the center of the array, the rest in the susceptible state.}
\label{patterns}
\end{figure}

\begin{figure}[ht]
\centering
\includegraphics[width=0.7\textwidth]{especNew.eps}
\caption{Power spectra of activity, from a single run, for the ordered phase, with $\sigma=2.5$ and $\sigma=5.5$. $K=6$ and $N=1000$.}
\label{fou}
\end{figure}

\begin{figure}[ht]
\centering
\includegraphics[width=0.6\textwidth]{newfig.eps}
\caption{At minimums parameter space is maximally {\it cuadriculado}. We obtain from our numerical experiments: $\triangleright/a=K+c$ (correlations proportional to coordination number). 
$numerito=(3\sigma_1-\sigma_3)^2+(2\sigma_1-\sigma_2)^2$, where $\sigma_1$ (used to estimate $\Delta_\sigma$) is the coupling at the transition in the disordered phase (see fig \ref{varsigma12}), 
$\sigma_2$ is the coupling for which $\varsigma=1$ in the disordered phase, and $\sigma_3$ is the coupling for which $\varsigma=1$ in the ordered phase (see fig \ref{deltasigma}). No minimum seems to exist for $K=4$.}
\label{new}
\end{figure}

\begin{figure}[ht]
\centering
\includegraphics[width=0.6\textwidth]{varsigma1.eps}
\caption{$\sigma_\varsigma(x)$ for $\varsigma\approx 1$ (continuous blue line), see white points in figure \ref{deltasigma}. 
We best fit to a stretched exponential (dashed red line), obtaining 
$\beta\approx0.84$ and $\Delta_\sigma\approx1.18$. We used a better method to obtain $\sigma_\varsigma(x)$ than in previous version.
The quantity $\Delta_\sigma$ is the difference in coupling $\sigma$ between the two plateaus.}
\label{var1}
\end{figure}

\begin{figure}[ht]
\centering
\includegraphics[width=0.9\textwidth]{figx3.eps}
\caption{Stretched exponential relaxation of activity $F$ along curves of constant $\varsigma$ ($F_\varsigma(x)$). Numerically obtained values
in continuous blue lines, best fit in dashed red lines.
The quantity $\Delta_F$ is the difference in activity between the two plateaus.}
\label{figx3}
\end{figure}

\begin{figure}[ht]
\centering
\includegraphics[width=0.9\textwidth]{figx4.eps}
\caption{$\Delta_\sigma(\varsigma)$, $\Delta_F(\varsigma)$, the exponents $\beta_\sigma(\varsigma)$
and $\beta_F(\varsigma)$ and, most important, $\beta_F(\Delta_F)$.}
\label{figx4}
\end{figure}

\begin{figure}[ht]
\centering
\includegraphics[width=0.9\textwidth]{profs.eps}
\caption{$F_\varsigma(z)$ for some $\varsigma<1$. At $\varsigma\sim0.82$ there's a {\it bump} in the profile, just before the decay in activity,
and at this value of $\varsigma$ $\beta\rightarrow 2$ (the gaussian; see the inset and figure \ref{figx4}).}
\label{profs}
\end{figure}

\end{document}